# Title

Lithium redistribution and electrode degradation in commercial NMC/Graphite and LCO/Graphite cells under high-rate cycling conditions

# Authors

Naisargi Kanabar[1] ( nkanabar@albany.edu ) - Corresponding Author
Nicholas Fusco [2] ( nafusco@albany.edu)
Seiichiro Higashiya[2] ( shigashiya@albany.edu )
Devendra Sadana[2] ( dsadana@albany.edu )
Stephen Bedell[2] ( sbedell@albany.edu )
Haralabos Efstathiadis[2] ( hefstathiadis@albany.edu )

# Affiliations

[1]Department of Physics, University at Albany (SUNY), Albany, NY 12222
[2]Department of Nanoscale Science and Engineering, University at Albany (SUNY), Albany, NY 12222

# Abstract

Fast charging accelerates degradation in lithium-ion batteries, but the effects of cathode chemistry and charging protocol remain difficult to separate. Here, commercial NMC811/graphite (200 mAh) and $LiCoO_2$/graphite (45 mAh) coin cells were cycled for 100 cycles at 0.3C, 1C, 2C, and 3C, followed by scanning electron microscopy (SEM), X-ray diffraction (XRD), and nuclear reaction analysis (NRA). Both cell types exhibited a transition in cycling stability between 1C and 2C. NMC811/graphite retained approximately 99.7% of its initial discharge capacity at 0.3C and 1C, decreasing to 72.3% at 2C and 63.5% at 3C, while $LiCoO_2$/graphite retained only 11.1% at 2C. NRA revealed maximum lithium accumulation in graphite anodes at 2C for both chemistries, supported by SEM observations of granular surface deposits. XRD showed rate-dependent shifts in cathode (003) and graphite (002) reflections, indicating structural changes during cycling. The lower lithium accumulation at 3C was associated with reduced charge throughput, while the $LiCoO_2$/graphite 3C condition was limited by its charging protocol. These findings highlight the importance of anode-side degradation and charging protocols in interpreting high-rate cycling performance of commercial lithium-ion batteries.

# Keywords

Fast charging; Capacity degradation; Nuclear reaction analysis; Lithium accumulation; Cathode chemistry; Charging protocols.

# 1. Introduction

Batteries underpin applications from portable electronics to electric vehicles, and lithium-ion chemistry dominates that space because of its high energy density. Performance under fast charging remains the limiting factor, however, and the rate at which a cell can be charged has become one of the principal constraints on deployment [1]. High currents reduce energy efficiency and accelerate the loss of both capacity and power [1,2]. The pursuit of higher capacity and energy density is further constrained by the positive electrode, which governs $Li^+$ transport, sets much of the achievable performance, and represents the largest single contribution to cell cost [3]. Considerable effort has gone into improving durability under aggressive cycling, yet understanding how different cathode chemistries respond to increasing C-rates in commercial full cells remains challenging. This question is particularly important for layered oxides, where differences in composition and structure can lead to distinct degradation pathways.

$LiCoO_2$ (LCO) was the first layered oxide to be identified as a practical intercalation host [4] and was used in the first commercial lithium-ion battery introduced by Sony in 1991 [5]. That milestone triggered the rapid growth of lithium-ion technology. More than four decades after the material was first described, and despite the emergence of lithium iron phosphate (LFP), lithium manganese oxide (LMO), nickel-manganese-cobalt (NMC) and nickel-cobalt-aluminium (NCA) alternatives, LCO remains in widespread commercial use [6]. Its persistence rests on a high theoretical capacity of 274 mAh $g^{-1}$ and a high crystallographic density of 5.1 g $cm^{-3}$, which together provide the potential for high volumetric energy density required in consumer electronics. Its theoretical specific energy is approximately 1070 Wh $kg^{-1}$ at an average voltage of 3.9 V. Raising the C-rate increases deliverable power but can shorten cycle life and substantially reduce accessible capacity, and efforts to improve its rate capability by surface and structural modification continue [7]. Concerns over cobalt, particularly its limited availability, high cost, and ethical sourcing, have nonetheless driven sustained effort toward cobalt-reduced and cobalt-free alternatives [8].

A route toward reducing cobalt content emerged in the late 1990s with $LiNi_{1-x-y}Co_xMn_yO_2$ [9], and the equiatomic composition $LiNi_{1/3}Mn_{1/3}Co_{1/3}O_2$ (NMC111) was established shortly afterwards [10]. NMC111 was widely adopted in early electric-vehicle applications, but its specific energy is insufficient for current targets, and its cobalt content remains costly. Raising the nickel fraction increases capacity, and the nickel-rich end of the series culminated in $LiNi_{0.8}Mn_{0.1}Co_{0.1}O_2$ (NMC811), which delivers roughly 200 mAh $g^{-1}$ at an average discharge voltage near 3.8 V and is now a leading candidate for high-energy-density applications [11].

The higher capacity of nickel-rich oxides comes at the cost of reduced structural and interfacial stability [11,12]. NMC811 degrades through several well-documented mechanisms, including microcracking of secondary particles, surface reconstruction to rock-salt phases, lattice oxygen release, and subsequent electrolyte reactions [13,14,15,16]. Researchers have investigated its structural evolution and lithium dynamics during cycling in detail [17,18]. In contrast, LCO generally exhibits relatively good structural stability when operated within its conventional voltage window of approximately 4.2 V, although its high-rate performance can still be limited by electrode kinetics, polarization, and cell design. At higher voltages, LCO can undergo structural transformations that accelerate degradation [19]. Based on cathode chemistry, NMC811 is expected to be more sensitive to high-rate cycling, whereas LCO cycled within its conventional voltage range is expected to exhibit greater structural stability. However, the high-rate performance of commercial lithium-ion cells depends not only on cathode chemistry but also on electrode design, charging protocols, and interfacial processes. Consequently, differences in capacity degradation between commercial cell chemistries cannot necessarily be attributed to cathode stability alone.

Degradation in a full cell is conventionally described in terms of loss of cyclable lithium inventory, loss of active material at either electrode [20]. These degradation modes are commonly investigated using electrochemical measurements, including coulombic efficiency, incremental-capacity analysis, and electrochemical impedance spectroscopy [21,22]. However, such diagnostics are indirect: they provide information about changes in electrochemical behavior without directly identifying the location or chemical state of lithium, and distinct degradation mechanisms can produce similar capacity trajectories [23]. At high charging rates, lithium plating on graphite is a particular concern because it can contribute to irreversible lithium inventory loss through the formation of electrically isolated lithium and additional interfacial reactions [24,25,26]. Ion beam methods can directly measure lithium in harvested electrodes, providing complementary information about lithium distribution that cannot be obtained from electrochemical measurements alone [27,28].

Rate-capability studies are usually reported in terms of C-rate, with the influence of charging protocols receiving comparatively less attention. During constant-current (CC) charging, increasing the C-rate produces greater polarization, causing the cell to reach its upper cutoff voltage at a lower state of charge. In constant-current–constant-voltage (CC–CV) charging, the subsequent CV step allows additional charge to be delivered, although its contribution depends on the charging rate, hold duration, and current termination conditions. At higher C-rates, the CV step can account for a substantial fraction of the total charge, and a fixed-duration hold may terminate charging before the cell reaches a comparable state of charge. Consequently, incomplete charging can reduce the accessible discharge capacity and complicate the distinction between protocol-related capacity limitations and irreversible degradation. This issue is particularly important for commercial full cells, where the individual electrode potentials cannot be directly measured without a reference electrode. Therefore, charging protocols must be considered when interpreting the rate-dependent degradation behavior of commercial lithium-ion cells.

No single post-mortem characterization technique can fully resolve the underlying degradation mechanisms, making a combination of complementary methods important. X-ray diffraction (XRD) provides information about crystalline structural changes, including variations in graphite interlayer spacing and lattice parameters of layered oxide cathodes. Nuclear reaction analysis (NRA) enables direct measurement of lithium in harvested electrodes and provides information about lithium distribution. Scanning electron microscopy (SEM) reveals changes in electrode morphology, including surface deposits, particle cracking, and structural damage. Combining these techniques with electrochemical measurements therefore provides a more comprehensive understanding of rate-dependent degradation, linking changes in capacity with electrode structure, morphology, and lithium distribution.

In this work, commercial NMC811/graphite and LCO/graphite coin cells were investigated at four C-rates (C/3, 1C, 2C, and 3C) over 100 cycles, using one fresh cell for each rate. For NMC811/graphite cells, constant-current–constant-voltage (CC–CV) charging was applied at C/3, while constant-current (CC) charging was used at 1C, 2C, and 3C. In contrast, LCO/graphite cells were charged using the CC–CV protocol at all four C-rates. Electrochemical cycling combined with post-mortem XRD, SEM, and NRA measurements to investigate capacity degradation, structural and morphological changes, and lithium redistribution after cycling. This combined approach enables the evaluation of rate-dependent degradation behavior in both commercial cell systems while considering the influence of their different charging protocols on the observed electrochemical performance.

# 2. Experimental

## 2.1 Battery samples

Commercial lithium-ion coin cells with nominal capacities of 200mAh LIR3032 (Cornell Dubilier Knowles, USA) and 45mAh LIR2032 (EEMB Battery, USA) were used in this study. The 200mAh cells consisted of nickel-rich manganese cobalt oxide (NMC) as the cathode and graphite as the anode, while the 45mAh cells consisted of lithium cobalt oxide (LCO) as the cathode and graphite as the anode.

## 2.2 Cell cycling protocol

Commercial NMC/graphite and LCO/graphite coin cells were cycled at room temperature under different C-rate conditions. For both cell chemistries, the 0.3C condition was cycled using a CC–CV protocol. At higher C-rates, different cycling protocols were employed: NMC/graphite cells were cycled using a CC protocol, whereas LCO/graphite cells were cycled using a CC–CV protocol.

For NMC/graphite cells, cycling was performed at 0.3C, 1C, 2C, and 3C for 100 cycles. The 0.3C cells were cycled between 3.0 and 4.2 V using CC–CV, whereas the 1C, 2C, and 3C cells were cycled between 2.7 and 4.3 V using CC, with the wider voltage window selected to impose accelerated degradation conditions. High-C-rate cycling of the NMC/graphite cells was performed using a Gamry Reference 600+. LCO/graphite cells were also cycled at 0.3C, 1C, 2C, and 3C for 100 cycles using a CC–CV protocol within a voltage window of 2.75–4.2 V. The 0.3C, 1C, and 2C tests were conducted using a PAT-Tester-i-16 (EL-CELL). Due to the current limitation of this system (~100 mA), the 3C test was performed using an Arbin BT2453.

Following cycling, the cells were disassembled in the discharged state inside an argon-filled glovebox (0.5 ppm $H_2O$ and 0.2 ppm $O_2$) to minimize air exposure. The recovered graphite anodes and cathodes were rinsed by soaking twice in dimethyl carbonate (DMC) for 10 min to remove residual electrolyte. The electrodes were subsequently dried inside the glovebox and sealed in airtight pouches prior to further characterization.

## 2.3 Physical characterization

Morphological analysis was performed using a TESCAN MIRA3 scanning electron microscope (SEM) equipped with a backscattered electron detector. X-ray diffraction (XRD) measurements were carried out using a Rigaku SmartLab SE multipurpose θ-θ diffractometer with Cu Kα radiation, equipped with a HyPix-400 two-dimensional hybrid pixel detector, at 20 °C. Unit-cell parameters were determined via Le Bail refinement of the powder diffraction patterns using SmartLab Studio II software (version 4.6.411.0). Lithium distribution in the cathode and anode samples was investigated by nuclear reaction analysis (NRA) using the $^{7}Li(p,\gamma)^{8}Be$ reaction. The NRA response was obtained as lithium counts (ROI integral) as a function of incident proton energy. Further details of the NRA measurement methodology and the energy-to-depth conversion procedure are provided in Naisargi et al. [28]

# 3. Results and Discussion

## 3.1 Cycling on commercial cell

The voltage–capacity profiles and capacity retention of both NMC/Gr and LCO/Gr cells showed a strong C-rate and cell cycling protocol dependence (Fig. 1 and Table 1). For NMC/Gr, the 0.3C and 1C cells remained highly stable after 100 cycles, retaining approximately 99.7% of their second-cycle discharge capacity. At higher rates, capacity retention decreased to 72.3% at 2C and 63.5% at 3C, indicating pronounced degradation above 1C. The largest absolute capacity loss occurred at 2C (44.99 mAh), despite the lower retention at 3C, due to the substantially lower accessible capacity at 3C.

LCO/Gr exhibited a stronger high-rate response. Capacity remained stable at 0.3C and showed only minor degradation at 1C, whereas the 2C cell retained only 11.1% after 100 cycles, corresponding to a capacity loss of 31.48 mAh. The 3C cell exhibited very low accessible capacity from the initial cycles; however, this condition was strongly influenced by the charging protocol and therefore was not considered representative of an intrinsic 3C degradation trend.

The influence of the charging protocol was also considered when interpreting the C-rate dependence. At 0.3C, both NMC/Gr and LCO/Gr cells were charged using CC–CV, whereas at higher rates NMC/Gr was charged using CC and LCO/Gr using CC–CV with a fixed 60-s CV hold. NMC/Gr exhibited substantially lower capacity loss than LCO/Gr at high rates; however, this difference cannot be attributed solely to the charging protocol because of the different cell chemistries. For LCO/Gr, the contribution of the CV step increased with C-rate as polarization increased. At 3C, substantial current remained at the end of the 60-s CV hold, indicating incomplete charging. Thus, the observed high-rate behavior reflects the combined influence of C-rate, cell chemistry, and charging protocol, particularly for the LCO/Gr 3C condition.

## 3.2 Evolution of discharge profiles with cycling

The evolution of cell voltage as a function of discharge capacity at cycles 2, 25, 50, 75, and 100 further demonstrates the strong C-rate dependence of both cell chemistries (Fig. 2). For NMC811/Gr, the C/3 and 1C discharge profiles remain nearly superimposed throughout 100 cycles, consistent with the high-capacity retention reported in Table 1. In contrast, the 2C and 3C cells exhibit progressive contraction of the discharge profiles and shifts toward lower cell voltage with cycling. The initial discharge voltages at cycle 2 were approximately 4.12 and 4.07 V at 2C and 3C, respectively, decreasing to approximately 4.06 and 3.96 V by cycle 100 (Table 2). These changes are consistent with an aging-induced increase in polarization and a progressive reduction in accessible capacity during high-rate cycling [29].

A stronger rate-dependent evolution is observed for LCO/Gr. At C/3 and 1C, the discharge profiles remain relatively stable throughout cycling, consistent with the corresponding capacity-retention values reported in Table 1. The initial discharge voltages at cycle 2 were approximately 4.18 and 4.14 V at C/3 and 1C, respectively, with little change during subsequent cycling. In contrast, the 2C cell exhibits substantial progressive shortening and downward shifting of the discharge profile. The initial discharge voltage at 2C decreases from approximately 4.13 V at cycle 2 to 4.06 V at cycle 100. This voltage shift, together with the severe capacity loss reported in Table 1, is consistent with increased polarization and degradation during high-rate cycling.

The LCO/Gr cell cycled at 3C exhibits a distinctly different response, characterized by very low accessible capacity from the initial cycles. The initial discharge voltage was approximately 3.50 V at cycle 2 and showed no change with subsequent cycling. However, this apparent voltage stability should not be interpreted as improved electrochemical stability. As discussed above, the fixed 60-s CV hold resulted in incomplete charging at 3C, restricting the accessible state-of-charge and capacity range. Therefore, the LCO/Gr 3C condition is considered protocol-limited and is not interpreted as representative of the intrinsic degradation behavior at 3C.

## 3.3 SEM data

For the LCO/Gr cell (Fig. 3), the graphite anode largely retained its characteristic graphite particle structure at 0.3C and 1C. The most pronounced surface change occurred at 2C (Fig. 3g), where the graphite flakes were almost completely covered by a rough, granular layer with visible cracks. Interestingly, the 3C anode remained relatively clean because protocol-limited cycling prevented the cell from fully charging and discharging, thereby limiting lithium transfer and surface-deposit formation compared with 2C. On the cathode side (Fig. 3a-d), the LCO particles showed increasingly pronounced bright features along the particle edges at higher C-rates. These changes indicate increased surface morphological heterogeneity at high C-rates.

For the NMC/Gr cell (Fig. 4), the graphite anodes at 0.3C and 1C also remained relatively clean, whereas the 2C anode showed extensive rough and porous surface deposits covering the graphite morphology. At 3C, the surface deposits were non-uniform, and regions with clearly visible graphite particles were still observed. In contrast to the anode, the NMC cathode maintained a similar secondary-particle and primary-crystallite morphology from the pristine condition through 3C, with no clear rate-dependent morphological degradation observed in the available top-view SEM images.

These SEM observations are consistent with the Li-NRA results, which showed the highest anode lithium accumulation at 2C in both NMC/Gr and LCO/Gr cells, corresponding to the strongest surface deposits observed by SEM. This agreement suggests that the increased lithium detected by NRA is associated with material accumulated at or near the graphite surface. However, SEM alone cannot determine whether these deposits are metallic lithium or lithium-containing interphase products.

## 3.4 XRD data

XRD analysis revealed distinct C-rate-dependent structural evolution in the NMC/Gr and LCO/Gr cells. For NMC (Fig. 5a), the (003) reflection progressively shifted toward lower 2θ from 18.66° at 0.3C to 18.64°, 18.61°, and 18.59° at 1C, 2C, and 3C, respectively. The corresponding increase in $d_{003}$ spacing which can be calculated by Bragg's law, and hence in the c-lattice parameter ($C = 3d_{003}$), indicates progressive expansion of the layered NMC structure along the c-axis with increasing C-rate [18]. In contrast, LCO (Fig. 6a) exhibited a non-monotonic response, with the (003) peak shifted from 18.84° at 0.3C to 18.79°, 18.92°, and 18.89° at 1C, 2C, and 3C, respectively. The corresponding $d_{003}$ values and c-lattice parameters indicate expansion at 1C, contraction at 2C, and slight re-expansion at 3C (Table 3).

The graphite anodes also showed chemistry-dependent structural behavior. For the NMC/Gr cell (Fig. 5b), the c-lattice parameters ($C = 2d_{002}$) were 6.73 Å at 0.3C, 6.72 Å at 1C, 6.73 Å at 2C, and 6.76 Å at 3C. Thus, the initial contraction at 1C was followed by recovery at 2C and more pronounced interlayer expansion at 3C. Conversely, the graphite anode in the LCO/Gr cell (Fig. 6b) exhibited c-lattice parameters of 6.73 Å at 0.3C, which decreased to 6.72 Å at 1C and remained nearly unchanged at 2C and 3C, indicating a modest overall contraction followed by structural stabilization at higher C-rates.

Overall, these results demonstrate distinct C-rate-dependent structural responses for the two cell chemistries. NMC undergoes progressive c-axis expansion with increasing C-rate, whereas LCO exhibits non-monotonic lattice evolution. Similarly, the graphite anode shows pronounced interlayer expansion at 3C in the NMC/Gr cell, while the graphite in the LCO/Gr cell exhibits a slight contraction at higher C-rates.

## 3.5 NRA data

The NRA measurements showed a clear variation in the graphite-anode lithium counts with cycling conditions. Using 0.3C as the reference condition (1.00) within each chemistry, the relative average counts at 1C, 2C, and 3C were 0.84, 20.27, and 1.15 for LCO/Gr, and 1.16, 8.16, and 2.08 for NMC/Gr, respectively. Thus, both chemistries showed the highest lithium accumulation at 2C. For NMC/Gr (Fig.7b), the lower signal at 3C compared with 2C can be associated with reduced accessible capacity and charge throughput, resulting in less lithium transfer during 100 cycles. For LCO/Gr (Fig.7d), this decrease was more pronounced because the 3C cycling was protocol-limited, resulting in incomplete charging and substantially reduced lithium transfer. Since the cycling protocols were not identical across all C-rates, these relative changes reflect the combined influence of C-rate and cycling protocols.

The cathodes exhibited a different trend. Relative to 0.3C (1.00), the NMC cathode (Fig. 7a) signal progressively decreased to 0.96, 0.93, and 0.91 at 1C, 2C, and 3C, respectively. In contrast, the LCO cathode (Fig. 7c) signal decreased to 0.95 at 1C and 0.83 at 2C, but increased slightly to 0.87 at 3C. The higher Li signal at 3C compared with 2C in LCO/Gr may be associated with the protocol-limited 3C cycling, which reduced the amount of lithium transferred during cycling. In addition, because the cells were disassembled after their final discharge at their respective C-rates, differences in cathode NRA signals may reflect both irreversible lithium loss and differences in the final state of lithiation caused by rate-dependent polarization [30].

The anode NRA results were consistent with the SEM observations, where the strongest graphite-surface deposits occurred at 2C, corresponding to the highest NRA lithium signals in both chemistries. This agreement suggests accumulation of lithium-containing material at or near the graphite surface. However, NRA and SEM alone cannot determine whether this material is metallic lithium, lithium-containing interphase species, or a combination of both.

# 4. Conclusions

Commercial NMC/Gr and LCO/Gr cells exhibited a clear transition in cycling stability between 1C and 2C. Both chemistries remained relatively stable at C/3 and 1C, whereas substantial capacity degradation occurred at higher rates. After 100 cycles, NMC/Gr retained 72.3% and 63.5% of its second-cycle capacity at 2C and 3C, respectively, while LCO/Gr retained only 11.1% at 2C. The largest absolute capacity loss occurred for NMC/Gr at 2C (44.99 mAh). The reduction in discharge capacity and voltage at higher C-rates further suggests increased polarization and reduced accessible capacity.

NRA and SEM revealed significant lithium accumulation and granular surface deposits on graphite anodes at 2C for both chemistries. Lithium accumulation increased by approximately 8.16 and 20.27 times relative to C/3 for NMC/Gr and LCO/Gr, respectively. The lower accumulation at 3C was associated with reduced charge throughput rather than improved stability. XRD showed rate-dependent structural changes but no detectable crystalline lithium-bearing phases on the anodes. The accumulated lithium-containing material may therefore be amorphous or nanocrystalline, although its chemical identity remains uncertain.

The charging protocol also influenced high-rate performance. For LCO/Gr, the contribution of the constant-voltage (CV) step increased substantially at higher C-rates. However, the fixed 60 s CV hold resulted in incomplete charging at 3C, making this condition protocol-limited rather than representative of intrinsic degradation behavior.

Overall, the combined electrochemical, NRA, SEM, and XRD results suggest that high-rate degradation is strongly associated with lithium accumulation and interfacial changes at the graphite anode. The observation of a similar 1C–2C transition in cells with different cathode chemistries, voltage windows, testing instruments, and charging protocols suggests a common rate-dependent limitation, although the influence of experimental conditions cannot be fully excluded. Future studies using identical CC and CC–CV protocols are needed to distinguish the effects of C-rate and charging conditions on capacity degradation and lithium redistribution.

# Figures

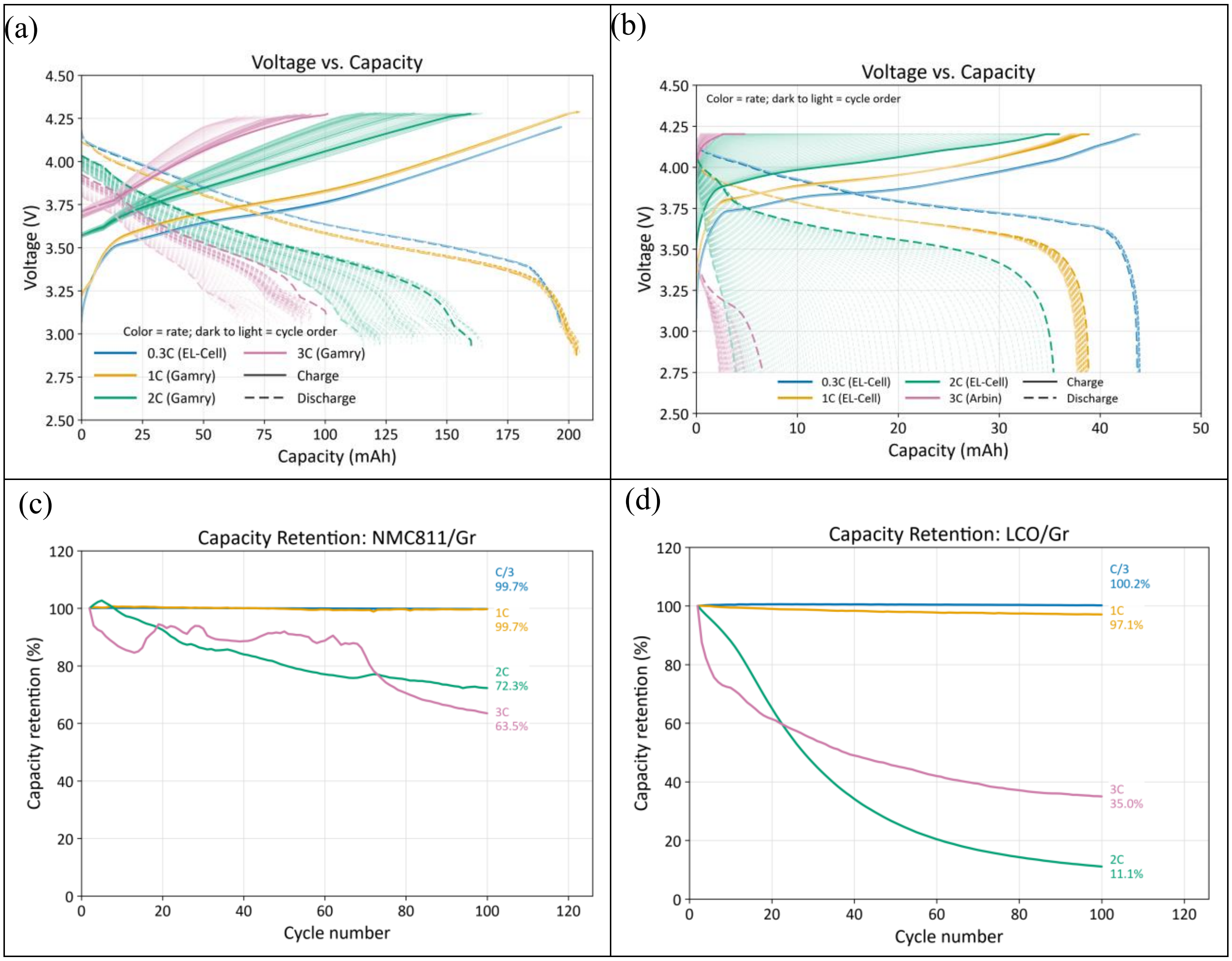

**Figure 1: Charge–discharge and Capacity retention profiles:** Commercial (a) NMC/Gr and (b) LCO/Gr cells cycled at different C-rates. Cell voltage is plotted as a function of charge/discharge capacity at 0.3C, 1C, 2C, and 3C. Capacity retention is a function of cycle number for (c) NMC/Gr and (d) LCO/Gr cells at the corresponding C-rates.

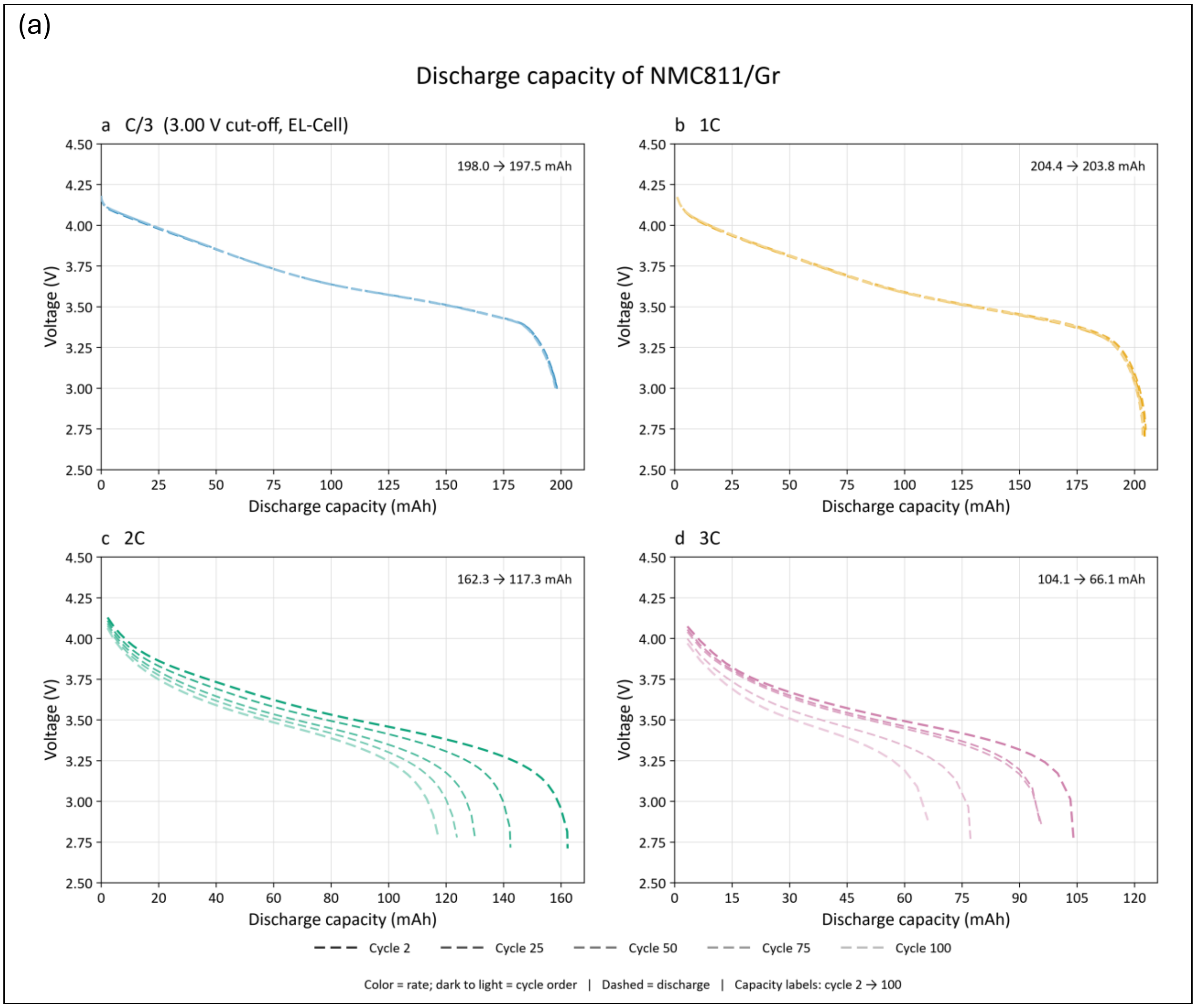
(a)
Discharge capacity of NMC811/Gr
a C/3 (3.00 V cut-off, EL-Cell)
198.0 → 197.5 mAh
b 1C
204.4 → 203.8 mAh
c 2C
162.3 → 117.3 mAh
d 3C
104.1 → 66.1 mAh
Voltage (V)
Discharge capacity (mAh)
Cycle 2
Cycle 25
Cycle 50
Cycle 75
Cycle 100
Color = rate; dark to light = cycle order | Dashed = discharge | Capacity labels: cycle 2 → 100

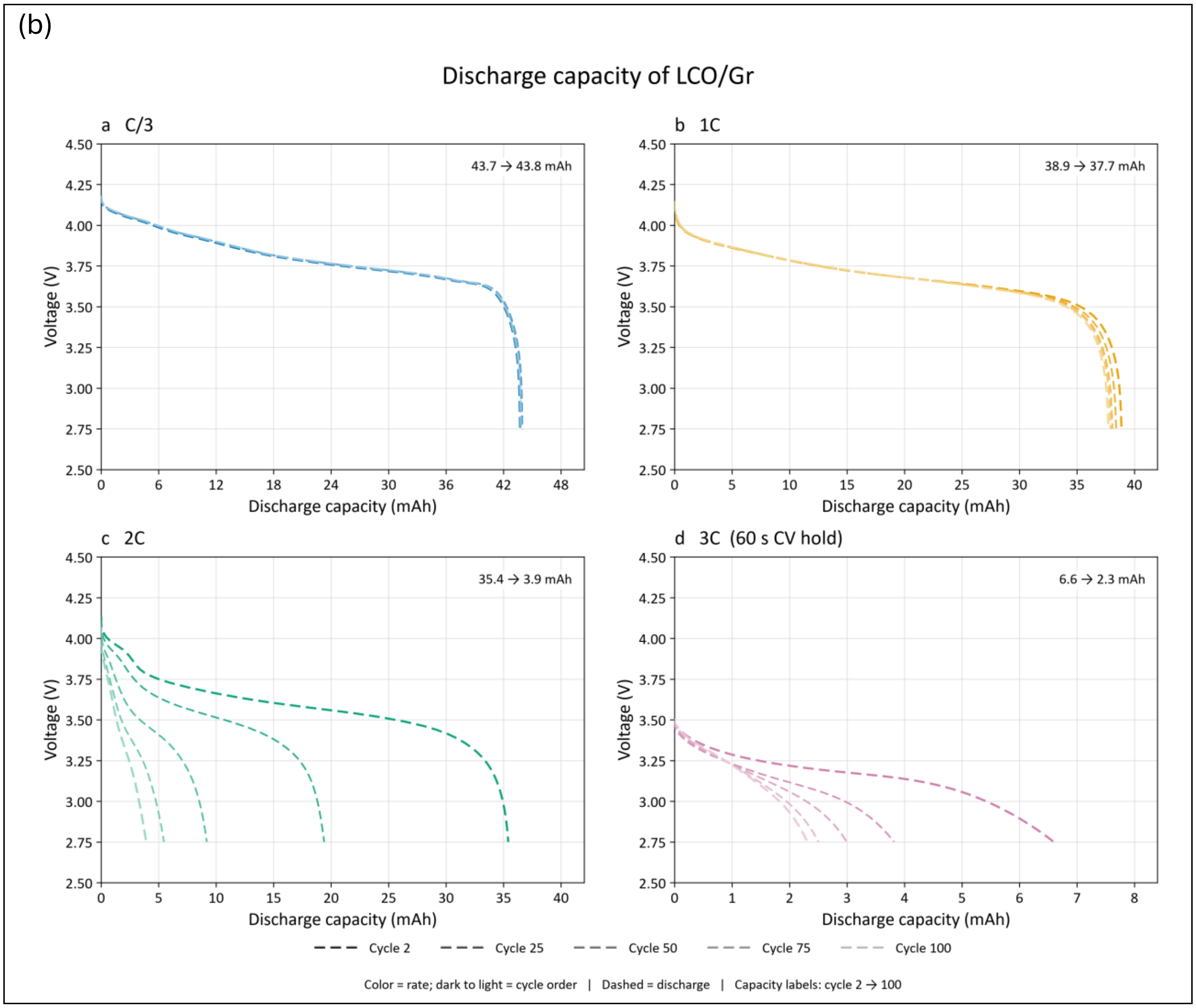

**Figure 2: Evolution of discharge voltage–capacity profiles:** (a) NMC/Gr and (b) LCO/Gr cells cycled at C/3, 1C, 2C, and 3C. Discharge profiles are shown at cycles 2, 25, 50, 75, and 100 to illustrate the effects of C-rate and cycling on cell voltage and accessible discharge capacity.

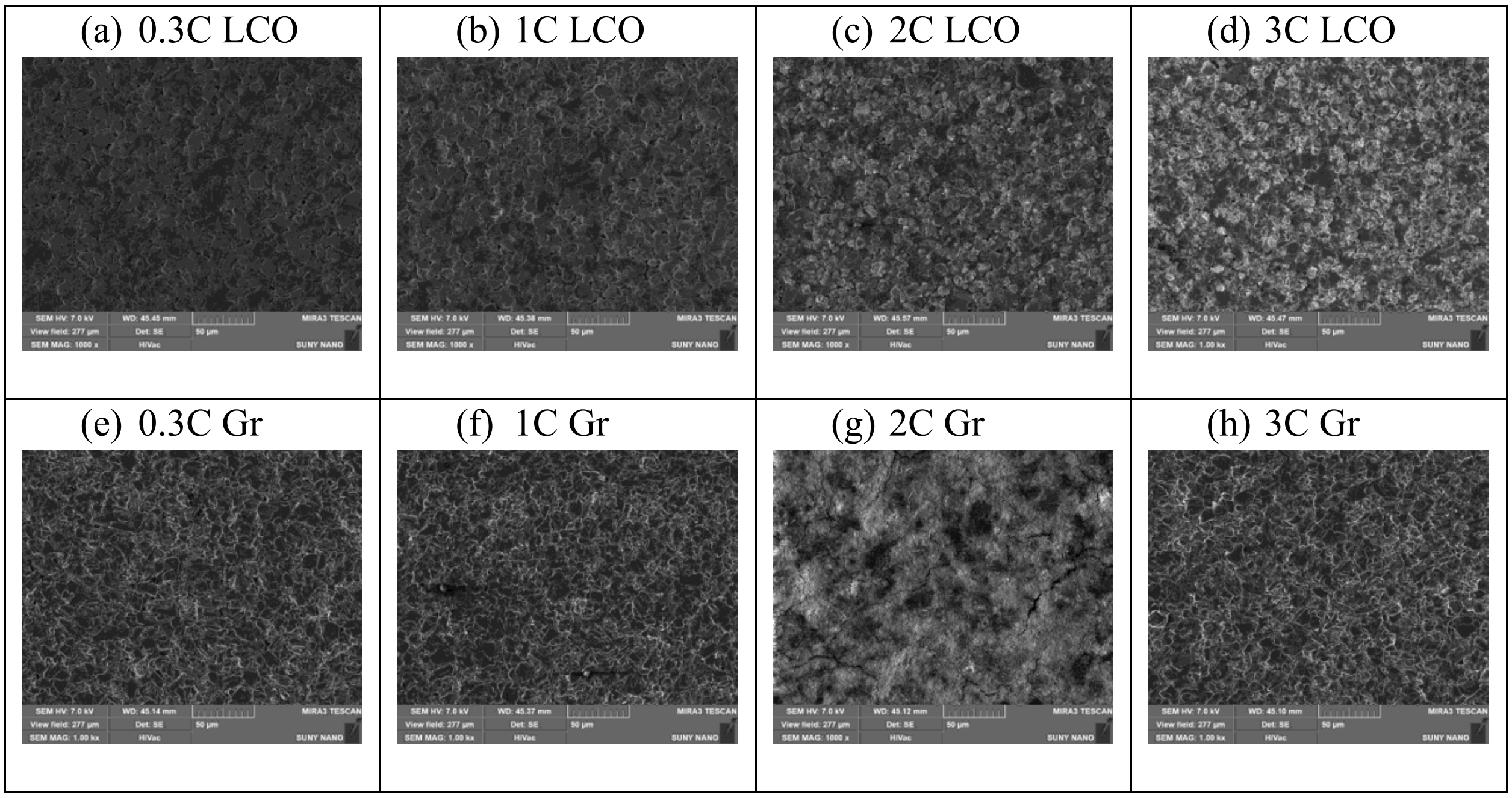

**Figure 3: SEM images:** (a–d) LCO cathodes and (e–h) graphite anodes recovered from LCO/Gr cells after 100 cycles at 0.3C, 1C, 2C, and 3C, acquired at 1000× magnification.

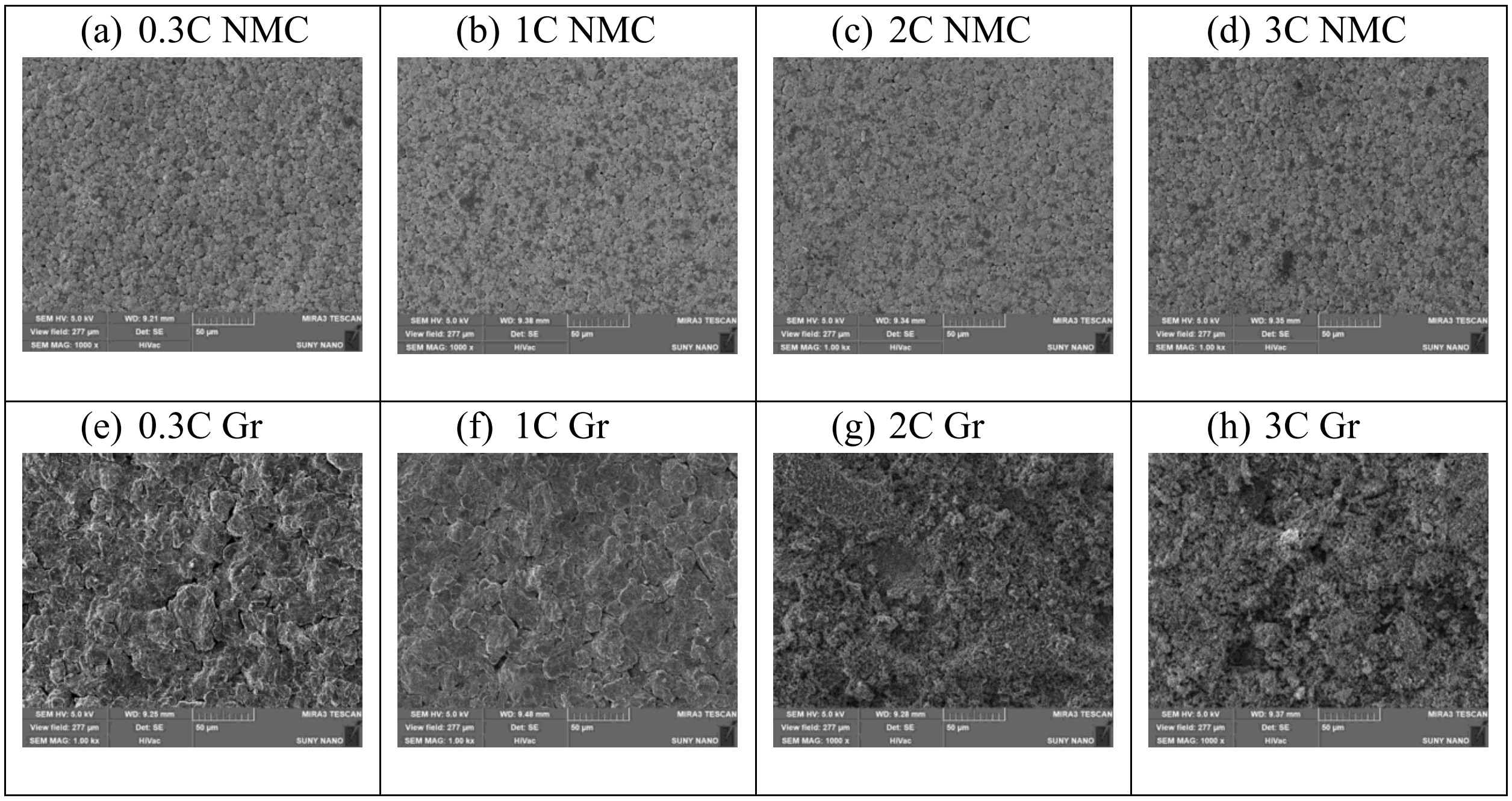

**Figure 4: SEM images:** (a–d) NMC cathodes and (e–h) graphite anodes recovered from NMC/Gr cells after 100 cycles at 0.3C, 1C, 2C, and 3C, acquired at 1000× magnification.

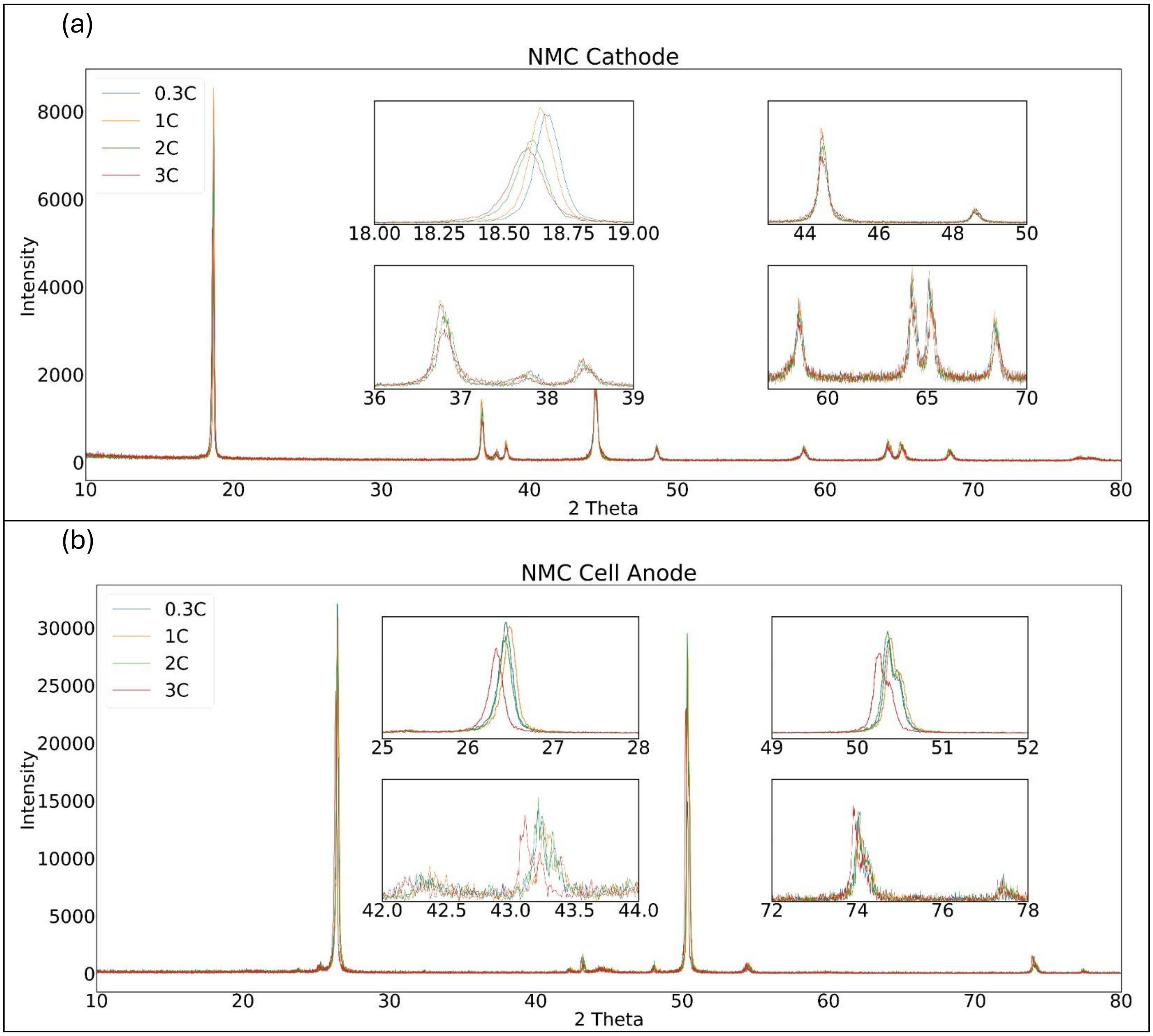

**Figure 5: XRD patterns:** (a) NMC811 cathodes and (b) corresponding graphite anodes after 100 cycles at different C-rates, showing a shift toward lower 2θ with increasing C-rate in NMC811 and graphite interlayer expansion at 3C.

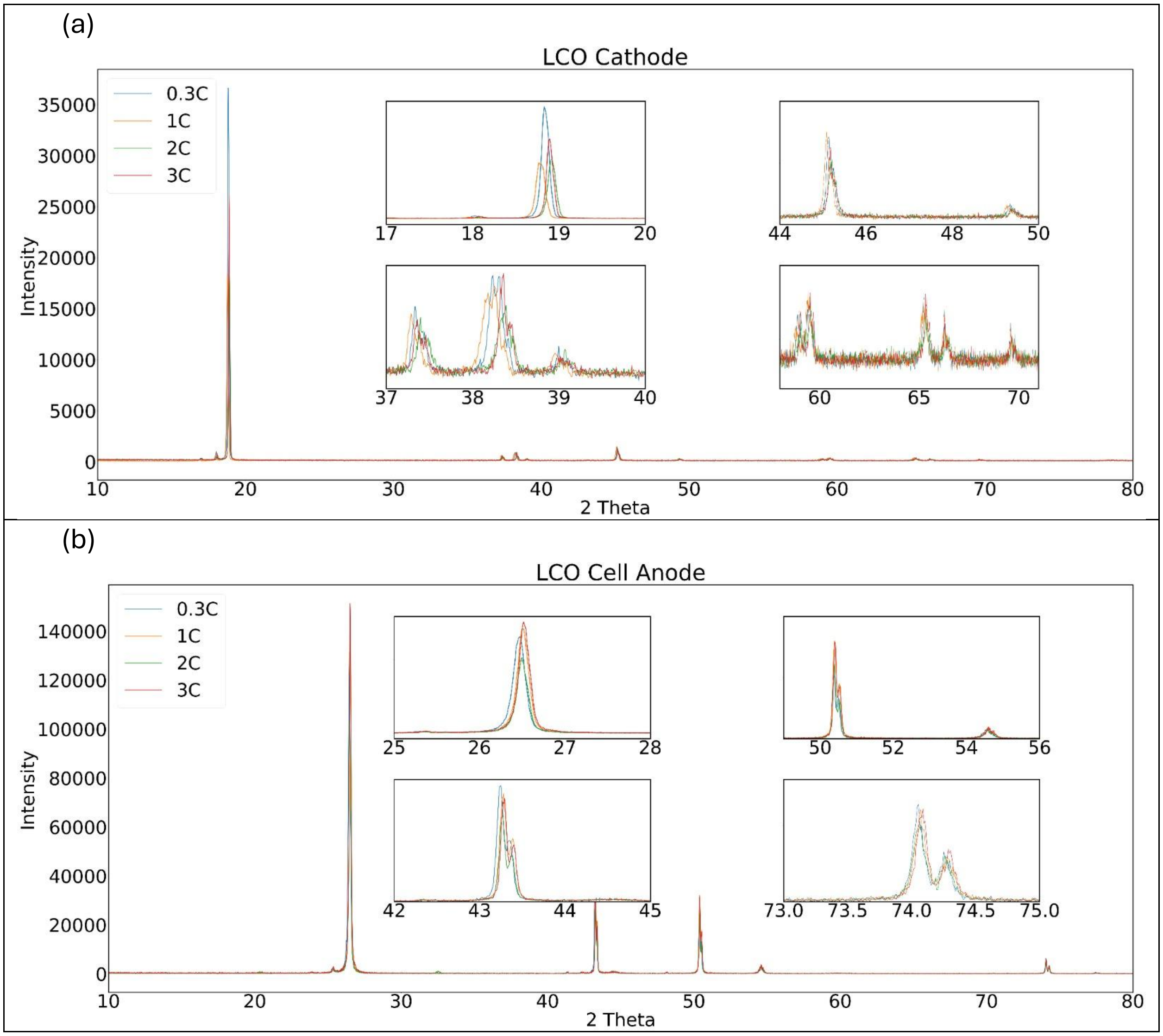

**Figure 6: XRD patterns:** (a) LCO cathodes and (b) corresponding graphite anodes after 100 cycles at different C-rates, showing non-monotonic peak shifts in LCO and rate-dependent changes in graphite interlayer spacing.

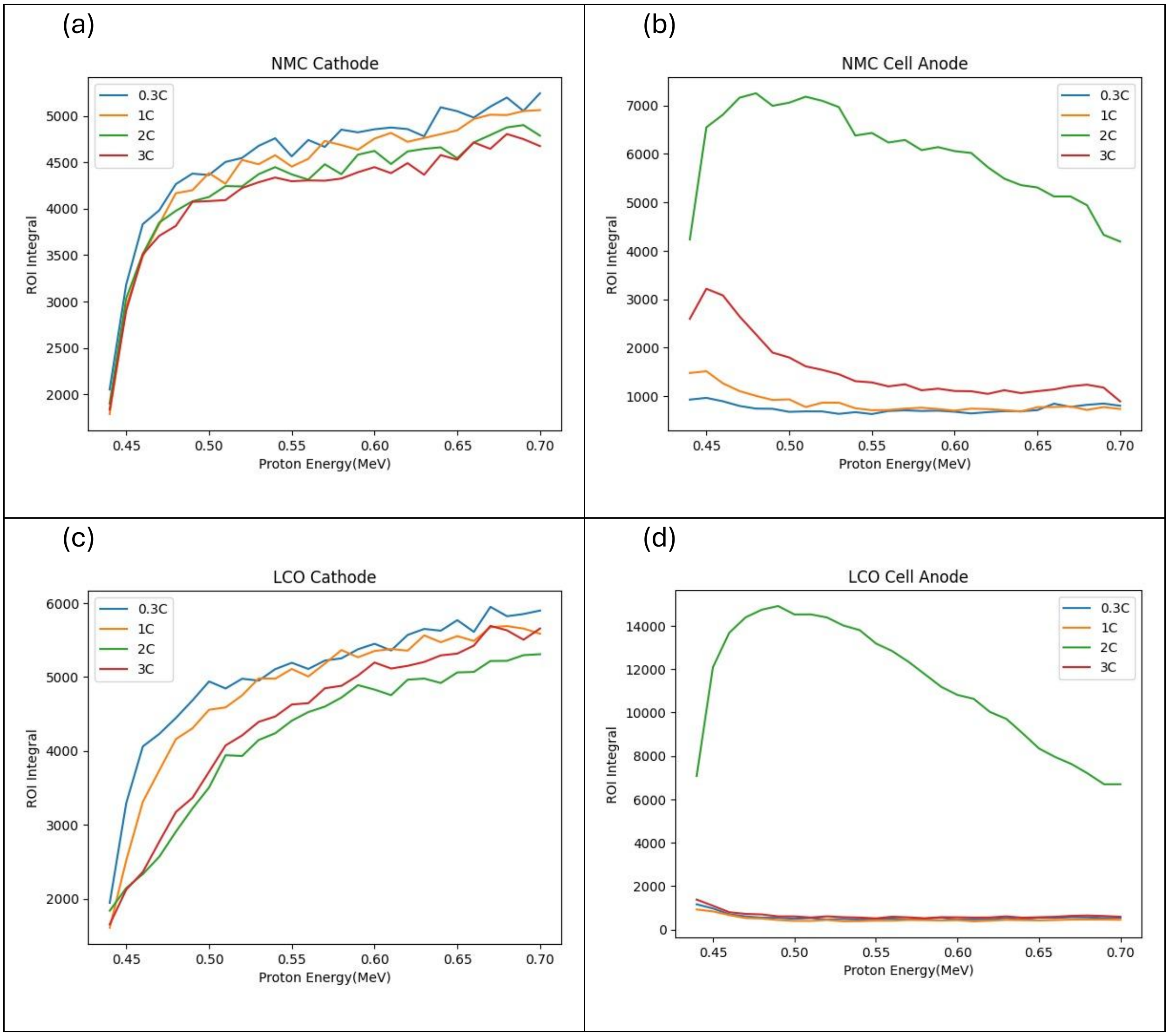

**Figure 7: NRA spectra:** It shows rate-dependent lithium counts in (a) NMC811 cathodes, (b) corresponding graphite anodes, (c) LCO cathodes, and (d) corresponding graphite anodes after 100 cycles at different C-rates (0.3C, 1C, 2C, and 3C).

# Tables

| Cell | C-rate | Discharge capacity, cycle 2 (mAh) | Discharge capacity, cycle 100 (mAh) | Capacity loss (mAh) | Retention (%) |
|---|---|---|---|---|---|
| NMC/Gr | 0.3C | 198.042 | 197.542 | 0.500 | 99.75 |
| NMC/Gr | 1C | 203.288 | 202.704 | 0.584 | 99.71 |
| NMC/Gr | 2C | 162.270 | 117.280 | 44.99 | 72.28 |
| NMC/Gr | 3C | 104.070 | 66.078 | 37.992 | 63.49 |
| LCO/Gr | 0.3C | 43.720 | 43.799 | −0.079 | 100.18 |
| LCO/Gr | 1C | 38.870 | 37.730 | 1.140 | 97.07 |
| LCO/Gr | 2C | 35.405 | 3.918 | 31.487 | 11.07 |
| LCO/Gr | 3C | 6.595 | 2.308 | 4.287 | 35.00 |

**Table 1.** Discharge capacity, capacity loss, and capacity retention of NMC/Gr and LCO/Gr cells cycled at different C-rates. Discharge capacities are reported for cycles 2 and 100, with capacity loss and retention calculated relative to cycle 2.

| Cell | C-rate | Current (mA) | Voltage cycle 2 (V) | Voltage cycle 100 (V) |
|---|---|---|---|---|
| NMC/Gr | C/3 | 66.7 | 4.1805 | 4.1807 |
| NMC/Gr | 1C | 200 | 4.172 | 4.1723 |
| NMC/Gr | 2C | 400 | 4.1282 | 4.0628 |
| NMC/Gr | 3C | 600 | 4.0732 | 3.9695 |
| LCO/Gr | C/3 | 15 | 4.1806 | 4.1826 |
| LCO/Gr | 1C | 45 | 4.1427 | 4.1491 |
| LCO/Gr | 2C | 90 | 4.1324 | 4.0681 |
| LCO/Gr | 3C | 135 | 3.5056 | 3.5056 |

**Table 2.** Initial discharge voltages at cycles 2 and 100 for NMC811/graphite and LCO/graphite cells cycled at different C-rates (0.3C, 1C, 2C, and 3C).

| Cell | C-rate | Cathode 2θ (003) | Cathode $d_{003}$ (Å) | Cathode $C = 3d_{003}$ (Å) | Anode 2θ (002) | Anode $d_{002}$ (Å) | Anode $C = 2d_{002}$ (Å) |
|---|---|---|---|---|---|---|---|
| NMC/Gr | C/3 | 18.658° | 4.7519 | 14.2557 | 26.443° | 3.3679 | 6.7359 |
| NMC/Gr | 1C | 18.640° | 4.7565 | 14.2694 | 26.488° | 3.3623 | 6.7246 |
| NMC/Gr | 2C | 18.613° | 4.7633 | 14.2899 | 26.443° | 3.3679 | 6.7359 |
| NMC/Gr | 3C | 18.595° | 4.7679 | 14.3036 | 26.335° | 3.3815 | 6.7630 |
| LCO/Gr | C/3 | 18.841° | 4.7061 | 14.1182 | 26.464° | 3.3653 | 6.7305 |
| LCO/Gr | 1C | 18.787° | 4.7195 | 14.1584 | 26.508° | 3.3598 | 6.7196 |
| LCO/Gr | 2C | 18.917° | 4.6872 | 14.0617 | 26.499° | 3.3608 | 6.7217 |
| LCO/Gr | 3C | 18.894° | 4.6929 | 14.0787 | 26.521° | 3.3581 | 6.7161 |

**Table 3.** XRD peak positions (2θ), interplanar spacings (d), and calculated lattice parameters (c) for the (003) cathode and (002) graphite anode reflections of NMC811/Gr and LCO/Gr cells after 100 cycles at different C-rates.

# 5. Acknowledgements

This research project was made possible by the generous support of the NYS Center for Advanced Technology in Nanomaterials and Nanoelectronics (CATN2). We are deeply grateful for their financial contribution and unwavering belief in our work, which has been instrumental in achieving our research goals. I am also thankful to Robert Bohl, Dr. Wei Zheng, and Dr. Daniele Cherniak, for their assistance with data collection.

# 6. Declaration of generative AI and AI-assisted technologies in the writing process

During the preparation of this work, the author(s) used ChatGPT in order to find and fix grammatical errors, voice dictation, and improve the readability of the manuscript. After using this tool/service, the author(s) reviewed and edited the content as needed and take(s) full responsibility for the content of the published article.